\magnification=\magstep1

%
\hsize = 6.50truein
\vsize = 8.50truein
\hoffset = 0.0truein
\voffset = 0.0truein
\lineskip = 2pt
\lineskiplimit = 2pt
\overfullrule = 0pt
\tolerance = 2000
\topskip = 0pt
\baselineskip = 18pt
\parindent = 0.4truein
\parskip = 0pt plus1pt
\def\medskip{\vskip6pt plus2pt minus2pt}
\def\bigskip{\vskip12pt plus4pt minus4pt}
\def\smallskip{\vskip3pt plus1pt minus1pt}
\centerline{\bf The NMR of High Temperature Superconductors}
\centerline{\bf without Anti-Ferromagnetic Spin Fluctuations}
\bigskip
\centerline{Jamil Tahir-Kheli}
\centerline{\it First Principles Research, Inc.}
\centerline{\it 8391 Beverly Blvd., Suite \#171, Los Angeles, CA 90048}
\centerline{\it www.firstprinciples.com}
\bigskip
\bigskip
\noindent{\bf Abstract} 
\bigskip
A microscopic theory for the NMR anomalies of the planar Cu and O sites
in superconducting
La$_{1.85}$Sr$_{0.15}$CuO$_4$ is presented that quantitatively explains
the observations without the need to invoke anti-ferromagnetic spin
fluctuations on the planar Cu sites and its significant discrepancy
with the observed incommensurate neutron spin fluctuations. 
The theory
is derived from the recently published ab-initio band structure
calculations that correct LDA computations tendency to overestimate
the self-coulomb repulsion for the half-filled Cu $d_{x^2-y^2}$ orbitals
for these ionic systems. The new band structure leads to two
bands at the Fermi level with holes
in the Cu $d_{z^2}$ and apical O $p_z$ orbitals in addition to the
standard Cu $d_{x^2-y^2}$ and planar O $p_\sigma$ orbitals. This band
structure is part of a new theory for the cuprates that explains
a broad range of experiments and is based upon the formation of
Cooper pairs comprised of a $k\uparrow$ electron from one band
and a $-k\downarrow$ electron from another band (Interband Pairing
Model).  
\vfill\eject
\noindent{\bf Introduction}
\medskip
All current explanations$^{1,2}$ of the dramatically different NMR behaviors
of the Cu and O nuclei separated by only $2.0$\AA\  in the CuO planes of high
temperature superconductors are based on the existence of
anti-ferromagnetic (AF) spin fluctuations on the Cu sites.  The NMR
difference
is attributed to the delicate cancellation on the O sites of these
AF fluctuations. The success of such models is dependent upon the 
AF spin correlation having a large peak very close or at wavevector
$(\pi,\pi)$.  Neutron scattering experiments have detected a spin
correlation peak at incommensurate wavevectors $(\pi\pm\delta,\pi)$
and $(\pi,\pi\pm\delta)$ with $\delta \approx 0.2\pi$, spoiling the
initial success of the models.$^{3}$

The models can be corrected by adding in next-nearest
neighbor hyperfine couplings of the Cu atoms to the O sites
to cancel the incommensurate fluctuations,$^{4}$ but the required 
hyperfine couplings are chemically too large.  Finally, these models
suffer from the lack of a microscopic derivation of the
wavevector, temperature, and doping dependence they require the
spin fluctuation function (i.e., the spin susceptibility
$\chi(q,kT)$) to satisfy in order to fit experiments. Thus, we regard
expressions for $\chi(q,kT)$ as empirically
devised to fit the NMR experimental data.

Recently, we proposed an Interband pairing model (IBP)$^{5,6}$ for
superconductivity that can explain the different Cu and O NMR without
invoking AF fluctuations and the functional form of the spin
susceptibility. In IBP, the incommensurate spin fluctuation peaks
observed by spin neutron scattering arise naturally from the
microscopic computed three dimensional (3D) band structure (but not
the 2D band structure we computed previously), yet they
do not lead to the NMR problems of AF spin fluctuation models.

The IBP model is based on the idea that in the vicinity of special
symmetry directions, Cooper pairs comprised of a $k\uparrow$ electron
from one band and a $-k\downarrow$ electron from a different band
are formed (interband pairs) and couple to standard BCS-like Cooper
pairs ($k\uparrow$ and $-k\downarrow$ from the same band) elsewhere
in the Brillouin zone. In particular for LaSrCuO, the crossing occurs
between a Cu $d_{x^2-y^2}$ band and a Cu $d_{z^2}$ band along the
diagonals $k_x=\pm k_y$.

Such a theory requires the existence of two bands at the Fermi level
that can cross with optimal doping associated with the bands
crossing at the Fermi level. Local Density Approximation (LDA) band
structure computations done over a decade ago$^{7}$ and accepted implicitly
by the physics community$^{8}$ as the correct starting point for developing
theories of the cuprates find only a single Cu
$d_{x^2-y^2}$ and O $p_\sigma$ antibonding band at the Fermi level
with all other bands well above or below the Fermi energy.
As we argued previously,$^{6}$ such calculations are plagued by improperly
subtracting only $(1/2)J$ for a half-filled band
from the $d_{x^2-y^2}$
orbital energy rather than subtracting a full $J$ due to the
strong correlation in such ionic systems,
where $J$ is the $d_{x^2-y^2}$ self-coulomb repulsion.
LDA calculations therefore,
artificially raise the $d_{x^2-y^2}$ and $p_\sigma$ antibonding
band above the other Cu $d$ bands.  When we correct the orbital energy
evaluation,$^{6}$ we find that in addition to the Cu $d_{x^2-y^2}$ and
O $p_\sigma$ orbitals, holes are created in the Cu $d_{z^2}$ and
apical O $p_z$ orbitals leading to two bands at the Fermi level.

The IBP model has had great success explaining a broad spectrum of
diverse high $T_c$ experimental observations.  These include the 
Hall effect, d-wave Josephson tunneling with coupling due to phonons,
the doping sensitivity of the cuprates, resistivity, and the NMR (within
the context of a 2D band structure with an approximate 3D structure
added on).$^{5}$
Most recently, we have shown$^{9}$ that our band crossing at
the Fermi level for optimal doping prevents the electron gas from
adequately screening the attractive electron-phonon coupling as
occurs in BCS superconductors, leading to a simple explanation for
the observed $T_c$ values in excess of standard BCS limits.  In addition,
the incommensurate neutron spin scattering and the
anomalous mid-infrared absorption peak arise in a straightforward
manner from our 3D bands.$^{10}$  Finally, the angle resolved photo-emission
spectroscopy (ARPES) and in particular, the observed so called
pseudo-gap (a gap on the Fermi surface in the normal state) for
underdoped cuprates and its disappearance for overdoping have been
explained as due to the rapid change in the orbital characters of the
two bands near the energy of the band crossing and the fact that $k$
states with primarily $d_{z^2}$ character do not have resolvable
quasiparticle peaks in the ARPES spectra.$^{11}$  This leads to the incorrect 
assignment of the Fermi surface in underdoped systems as the 
crossover surface between dominant $d_{x^2-y^2}$ and $d_{z^2}$
characters and hence the erroneous conclusion that a pseudo-gap
has opened above $T_c$ on the Fermi surface.  The reason for the lack
of a sharp quasiparticle peak in the ARPES spectra for $d_{z^2}$
electrons is because there is no great anisotropy in its dispersion
in the CuO planes versus its dispersion normal to the planes.
This is in contrast to the almost 2D dispersion of $d_{x^2-y^2}$
and leads to a large linewidth of $d_{z^2}$ from the intermediate
excited photo-electron state that is added to the physically
interesting linewidth of the initial electron state.

In this paper, we derive the key NMR observations based upon the detailed
3D band structure we obtained recently.$^{12}$  This explanation
supersedes the NMR discussion in our previous work that was
based upon a 2D band with an approximate 3D dispersion and is significantly
different in the details. The essential features remain the same.
These are: 

\smallskip\noindent
\item{1.)} the interesting region of the Brillouin zone (BZ) for the 
upper band is the vicinity of the saddle point density of states (DOS)
at $(\pi/a,0)$ or $(\pi/a,0,\pi/c)$ for the 3D zone and the 
relevant region for the lower band is near $(\pi/a,\pi/a)$ that
is at the top of this band.  Both the saddle point and the top
of the lower band at $(\pi/a,\pi/a)$ are close to the Fermi energy
(less than or on the order of $\approx 0.08$ eV).

\item{2.)} the character of the lower band $k$ states near 
$(\pi/a,\pi/a)$ has reduced O $p_\sigma$ and $2s$ character.  The
reduced O $p_\sigma$ is due to the O sites
forming a bonding combination in order to couple to $d_{z^2}$ at
$(\pi/a,\pi/a)$.  O $2s$ is reduced because it cannot couple to
$d_{z^2}$ and $d_{x^2-y^2}$ by symmetry.

\item{3.)} the Cu spin relaxation anisotropy of $\approx 3.4$ for
magnetic fields in the plane versus perpendicular to the CuO planes
is due to the small amount of Cu $d_{xy}$ and its spin orbital
coupling to $d_{x^2-y^2}$.

A new piece of chemistry appears in this paper in order to produce
the small increase ($\approx 0.1\%$) in the O spin relaxation rate
over temperature $(1/T_1 T)$ from 50K to 300K
that was not required in the 2D model.
That is the Jahn-Teller $5^\circ$ alternating tilt of the CuO$_6$
octahedra reducing the crystal point group from $D_{4h}$ to $D_{2h}$
and changing the Bravais lattice from body-centered tetragonal to
one-face-centered orthorhombic.$^{7}$  The distortion splits the saddle
point peak in the DOS at wavevector $(\pi/a,0,\pi/c)$ and 
$(0,\pi/a,\pi/c)$ into two peaks.  Hume-Rothery and Jahn-Teller
type arguments suggest the material will self-adjust to place its
Fermi level between these two peaks because the unperturbed saddle
point singularity is so close to the Fermi energy.
We argue, but do not compute,
that the distortion leads to the O $1/T_1 T$ increase with $T$ and
suggest this is the reason these systems have a tendency to self-dope
to optimal doping for the highest $T_c$.
\bigskip
\noindent{\bf 3D Band Structure}
\medskip
The 3D Fermi surface for optimally doped
La$_{1.85}$Sr$_{0.15}$CuO$_4$
is shown in figures $1(a-d)$ at various fixed $k_z$ values.$^{12}$
The range of
values of $k_z$ is $-2\pi/c<k_z<2\pi/c$ and $k_x$, $k_y$ vary
between $-\pi/a$ and $\pi/a$ 
where $a=3.8$\AA\ and 
$c=13.2$\AA\ are the doubled rectangular unit cell parameters.
As $k_z$ is increased from $0$ to $\approx 1.18\pi/c$,
the
Fermi surface is very similar to the standard LDA one band result
with a hole-like surface centered around $(\pi/a,\pi/a)$.  At
$k_z\approx 1.18\pi/c$, the top of the lower band is reached
and as $k_z$ is increased further, a second hole-like Fermi surface
appears centered around $(0,0)$.  This arises from the 3D coupling
of the apical O $p_z$ orbitals in one layer to its neighboring
apical O $p_z$ in another layer.  At $k_z\approx 1.54\pi/c$,
the two Fermi
surfaces touch along the diagonal that is the only symmetry allowed
crossing. Further increasing of $k_z$ to its upper limit of $2\pi/c$
splits the two surfaces into three with 
a hole-like surface centered around the
diagonal and two electron-like 
surfaces centered at $(\pi/a,0)$ and $(0,\pi/a)$.

Figures $2(a-d)$ show the total density of state (DOS) and the bare DOS for
Cu $d_{z^2}$, $d_{x^2-y^2}$ and O $p_\sigma$.  Note the large DOS
peak just below the Fermi level due to the almost pure 2D character
of the bands at $(\pi/a,\pi/a)$. At $(\pi/a,\pi/a)$, the lower band
is composed of $d_{z^2}$ and the bonding combination of $p_\sigma$.
This is the most unstable $d_{z^2}$ state at this $k$ vector.
There is no $d_{x^2-y^2}$ or O 2s due to symmetry. Because the
$p_\sigma$ orbitals are in a stabilizing
bonding combination, the most unstable $d_{z^2}$ state will 
not have much $p_\sigma$ character at all. Thus, the k states
that contribute to the peak in
the DOS just below the Fermi level in the vicinity of $(\pi/a,\pi/a)$
have very little $d_{x^2-y^2}$, $p_\sigma$, and O 2s characters.

The bare DOS for a given band is
defined as the product of the average orbital character 
for the band at a given energy times the DOS of the band.  The total
bare DOS for each orbital is the sum of the bare DOS from each band
and is the relevant quantity for the NMR.  Physically,
the total bare DOS of an orbital is the number of electrons in
the orbital per unit of energy.

The most important thing to notice from these figures is the
difference in the behaviors of the Cu $d_{z^2}$ bare DOS versus
$d_{x^2-y^2}$ and O $p_\sigma$.  $d_{z^2}$ is very sensitive
to the large DOS that arises from the lower band near $(\pi/a,\pi/a)$
whereas, both the total bare DOS for $d_{x^2-y^2}$ and $p_\sigma$
are not.  This is the fundamental reason for the difference in
the Cu and O spin relaxation rates.

Figures $3(a-c)$ show the orbital character for each band.  One can
see that the two bands trade off their orbital characters when
the bands cross.
\bigskip
\noindent{\bf The Cu and O NMR}
\medskip
We use standard expressions for the nuclear spin relaxation rates due
to delocalized electrons that are well described by simple Bloch
states to form bands.$^{13,14}$ All of the relevant expressions for
computing the spin relaxation rates on the planar Cu and O sites due
to $d_{x^2-y^2}$, $d_{z^2}$ and $p_\sigma$ are explicitly written down
in reference 5.  We do not reproduce them all here.  Instead, we will
write down the general form of the expression (equation (61) in
the above reference) to clarify our discussion of the results.

The general expression for the spin relaxation rate in the cuprates
where we neglect the contribution from the Cu 4s and O 2s contact
terms and the core polarization is,

$${1\over T_1}=2\biggl({2\pi\over\hbar}\biggr)(\gamma_e\gamma_h\hbar)^2
\int{\rm d}\epsilon
f(\epsilon)(1-f(\epsilon))\biggl\langle{1\over r^3}\biggr\rangle^2
[W_{\rm dip}(\epsilon) + W_{\rm orb}(\epsilon)],\eqno(1)$$

\noindent where $f(\epsilon)$ is the Fermi-Dirac function, 
$f(\epsilon)=1/(e^{\beta(\epsilon-\mu)} + 1)$
at energy $\epsilon$ and
$\mu$ is the chemical potential.  $\gamma_e$ and $\gamma_n$ are the
electronic and nuclear gyromagnetic ratios, $<1/r^3>$ is the mean
value of $1/r^3$ for the relevant orbital, $W_{\rm dip}(\epsilon)$ is
a function of the bare density of states of the orbitals for 
dipolar relaxation, and $W_{\rm orb}(\epsilon)$ is the similar expression
for orbital relaxation.

For Cu relaxation with the magnetic field normal to the CuO planes
(z-axis), $W_{\rm dip}$ and $W_{\rm orb}$ are given by,

$$W_{\rm dip}^z(\epsilon)=\biggl({1\over7^2}\biggr)
[6N_{d_{x^2-y^2}}(\epsilon)N_{d_{z^2}}(\epsilon) + 
N_{d_{x^2-y^2}}(\epsilon)N_{d_{x^2-y^2}}(\epsilon) +
N_{d_{z^2}}(\epsilon)N_{d_{z^2}}(\epsilon)
],\eqno(2)$$
$$W_{\rm orb}^z(\epsilon)=0,\eqno(3)$$

\noindent where $N(d_{x^2-y^2})(\epsilon)$ and
$N(d_{z^2})(\epsilon)$ are the total bare density of states for their
respective orbitals. We take $<1/r^3>=6.3$ a.u. for Cu$^{15}$ and
make the crude approximation$^{5}$ of $3.0$ a.u. for O.

The inclusion of Cu 4s and the effects of core polarization will lead
to a small change in the computed magnitude of the Cu spin relaxation
and Knight shift,
but should not change the overall qualitative behavior. The O 2s
can increase the magnitude of the O relaxation rate by an order of
magnitude due to its large density at the O nucleus but as discussed
above, cannot alter the qualitative behavior of the relaxation
and Knight shift curve because by symmetry, no 2s character appears
at $(\pi/a,\pi/a)$.

In most metals, the bare densities of states that appear in equations
(2) and (3) can be taken to be constant over the range $\mu\pm kT$
around the Fermi level. The integral in equation (1) is thereby
over $f(1-f)$ and is equal to the temperature $kT$. Hence,
$1/T_1 T$ is a constant.  Due to the band crossing, the 
closeness of the Fermi level to the saddle point singularity in the DOS at
$(\pi/a,0,\pi/c)$ and the top of the lower band,
the bare densities of states cannot be taken to be constant over
the range of energies relevant for computing the NMR. In addition,
the chemical potential $\mu$
increases with increasing temperature in order
to maintain particle conservation. Thus, $\mu$ must be solved for 
self-consistently at every temperature.

Figures 4a and 4b show the calculated Cu and O spin relaxation rates
over temperature $1/T_1 T$ for a z-axis magnetic field. The Cu
$1/T_1 T$ initially rises due to the sharp increase in the $d_{z^2}$
bare DOS just below the Fermi level from the DOS peak at $(\pi/a,\pi/a)$.
As the temperature is further increased, the chemical potential
increases to maintain particle conservation and the integral in equation
(1) ``falls over'' the top of the lower band leading to the sharp
decrease in the relaxation. The values we obtained are approximately
a factor of $2$ larger than experiment but the most important
point is the percentage increase from $50K$ to the maximum
and the approximately factor of $1.4$ decrease from the maximum
value to the value at $300K$ are compatible with experiment
where the increase is $\approx 10-30\%$ and the decrease is
about factor of $2$.$^{3}$

In contrast, the O $1/T_1 T$ decreases by $\approx 8\%$ as the
temperature is increased. This is due to the lack of the $(\pi/a,\pi/a)$
peak
in the bare DOS for $p_\sigma$ and the decrease in its bare DOS as
the energy is increased above the $T=0$ Fermi level. The relaxation 
is more sensitive to bare DOS values above the $T=0$ Fermi level
due to the increase in $\mu$. These numbers are about a factor of $5$
smaller than experimental values.  Inclusion of O 2s character can
easily produce an increase of a factor of $5-10$ without changing
the qualitative behavior.

The most important point to note here
is that the decrease of the O relaxation is very small compared to
the scale of the Cu relaxation decrease. Although, with the present
calculations the small observed increase is not reproduced, we have
already attained considerable success in obtaining such a dramatic
difference in the Cu and O NMR.  By considering the orthorhombic
CuO$_6$ tilt in the following section, we will argue that in fact,
the observed increase can be obtained by our model.

The expressions for the various Knight Shifts are explicitly written
down in reference 5 and are not reproduced here. As before$^{5}$, we
must assume that $d_{z^2}$ and Cu 4s interfere such that the
net dipolar field due to the $d_{z^2}$ and the Cu 4s hybrid is of the
opposite sign of a single $d_{z^2}$ in order to lead to an
increase in the Cu Knight shift with increasing temperature and
the lack of strong temperature dependence of the shift for a z-axis
field. This is discussed in detail in reference 5.

The one additional point in favor of the sign flip of the dipolar
field of $d_{z^2}$ due to interference with the 4s for our 3D
model as compared to our 2D model is that in the 3D model, 
$d_{z^2}$ holes appear in the vicinity of $(0,0,\pi/c)$.  At $(0,0)$,
4s character will mix with $d_{z^2}$ and by symmetry $p_\sigma$
cannot couple to them. Thus, one expects the 4s to mix into
the $d_{z^2}$ to increase the size of the $d$ orbital in the
planar directions, or in other words, interfere with $d_{z^2}$ with
the correct sign to lead to a net sign flip of the dipolar field.

Figures 5a and 5b show the Cu and O z-axis Knight shifts. The O
Knight shift does not include O 2s and
decreases with increasing temperature. The Cu
shift increases with increasing temperature and hence does not
track the spin relaxation curve in agreement with experiment.  This
is due to the fact that for relaxation, the DOS appear twice (squared)
in the relaxation expression because the initial and final state
probabilities of the relaxing electron must be multiplied together.
For the Knight shift, only a single power of the DOS appears. The
Cu relaxation is therefore more sensitive than the shift to the
sharp increase in the DOS due to $(\pi/a,\pi/a)$ just below the
Fermi level.

Note also the contribution to the temperature dependence of the Cu
shift from $d_{x^2-y^2}$ is much smaller than the contribution
from $d_{z^2}$.  In figure $5a$, we plot the $d_{x^2-y^2}$
contribution multiplied
by $10$ and minus the $d_{z^2}$ shift to incorporate the
sign flip interference from Cu 4s.  The scale of the Cu shift is
consistent with with experiments.

\bigskip
\noindent{\bf Orthorhombic Distortion}
\medskip
The orthorhombic CuO$_6$ octahedra tilt in
La$_{1.85}$Sr$_{0.15}$CuO$_4$ splits the DOS peak at $(\pi/a,0,\pi/c)$
and $(0,\pi/a,\pi/c)$ into two peaks at energy shifts
$\epsilon_0\pm\delta$ where $\epsilon_0$ is the original energy.
This leads to a local gap in the energy between these two values
in the vicinity of the saddle point.  As the contribution to the
DOS is large here, one expects the total DOS to be much smaller
between the two peaks. Chemically, one expects the size of $\delta$ to
be on the order of $0.01-0.05$ eV or greater. The Fermi level will
therefore fall between the two peaks.  The overall effect on the 
Cu NMR will be small due to the dominance of the $(\pi/a,\pi/a)$ peak
for Cu.  On the other hand, this distortion will dramatically
change the O NMR from slightly decreasing to slightly increasing
as observed by experiment.  We believe this is the reason for
the O NMR increase with temperature for LaSrCuO.

One also expects the system will adjust itself to place
its Fermi level between the two peaks in order to lower its total
free energy. This is essentially a Hume-Rothery or Jahn-Teller
type argument.  Such a mechanism provides a simple explanation for
the tendency of several cuprates to ``self-dope'' to the optimal
doping for $T_c$. 
\bigskip
\noindent{\bf Conclusions}
\medskip
We have presented a theory for the NMR of LaSrCuO that explains
the observed different Cu and O relaxations as arising from 
two bands with Cu $d_{x^2-y^2}$, $d_{z^2}$, planar O $p_\sigma$, and
apical O $p_z$ characters.  These bands were derived ab initio$^{6,12}$
by correcting the improper accounting of the self-coulomb contribution
to the orbital energy in LDA band structure calculations.  The
theory resolves the NMR anomalies with a microscopic picture that
does not require the introduction of anti-ferromagnetic spin fluctuations
and
it's corresponding disagreement with the observed 
incommensurate neutron spin fluctuations.

The splitting of the saddle point singularity in the
density of states at $k$ vector 
$(\pi/a,0,\pi/c)$ and $(0,\pi/a,\pi/c)$ by the CuO$_6$ orthorhombic
distortion changes the O NMR from monotonically decreasing with
increasing temperature to monotonic increasing with temperature
by splitting the peak into two peaks.
\bigskip
\noindent{\bf Acknowledgments}
\medskip
We wish to thank Jason K. Perry with whom all parts of
this work was discussed.  We also wish to thank William A. Goddard III
for his insight and
encouragement during the development of the ideas presented
here and in previous publications.
\bigskip
\noindent{\bf REFERENCES}\smallskip
\item {1.} F. Mila and T.M. Rice, {\it Physica C} {\bf 157}, 561 (1989)
\item {2.} D. Pines, {\it Physica C} {\bf 282}, 273 (1997)
\item {3.} R.E. Walstedt, B.S. Shastry, and S-W. Cheong,
{\it Phys. Rev. Lett.} {\bf 72}, 3610
(1994)
\item {4.} Y. Zha, V. Barzkin, and D. Pines,
{\it Phys. Rev. B} {\bf 54}, 7561 (1996)
\item {5.} J. Tahir-Kheli, {\it Phys. Rev. B} {\bf 58}, 12307 (1998)
\item {6.} J.K. Perry and J. Tahir-Kheli, {\it Phys. Rev. B} {\bf 58},
12323 (1998); J.K. Perry, {\it J. Phys. Chem.} (submitted),
xxx.lanl.gov/abs/cond-mat/9903088,
www.firstprinciples.com
\item {7.} W.E. Pickett, {\it Rev. Mod. Phys.} {\bf 61}, 433 (1989) and
references therein
\item {8.} P.W. Anderson, {\it The Theory of Superconductivity in
the High-T$_c$ Cuprates} (Princeton, 1997), p. 33
\item {9.} J. Tahir-Kheli, {\it Phys. Rev. Lett.} (submitted),
www.firstprinciples.com
\item {10.} J. Tahir-Kheli, (to be published)
\item {11.} J.K. Perry and J. Tahir-Kheli,
{\it Phys. Rev. Lett.} (submitted),
xxx.lanl.gov/abs/cond-mat/9908308,
www.firstprinciples.com
\item {12.} J.K. Perry and J. Tahir-Kheli,
{\it Phys. Rev. Lett.} (submitted),
xxx.lanl.gov/abs/cond-mat/9907332,
www.firstprinciples.com
\item {13.} C.P. Slichter, {\it Principles of Magnetic Resonance Third
Edition} (Springer-Verlag, 1990)
\item {14.} A. Abragam, {\it Principles of Nuclear Magnetism} (Oxford,
1961)
\item {15.} A. Abragam and B. Bleaney, {\it Electron Paramagnetic
Resonance of Transition Ions} (Dover, 1986), p. 458
\vfil\eject
\noindent{\bf Figure Captions}\bigskip
\item {1(a-d).} The Fermi surface for a.) $k_z=0$, b.) $k_z=1.30\pi/c$,
c.) $k_z=1.54\pi/c$, and d.) $k_z=2\pi/c$.
\item {2(a-d).} The density of states of the two bands and the bare
density of states for Cu $d_{x^2-y^2}$, $d_{z^2}$, and O $p_\sigma$
 in units 1/(eV$\times$spin$\times$unit cell).
\item {3(a-c).} The orbital characters for the 
Cu $d_{x^2-y^2}$, $d_{z^2}$, and O $p_\sigma$ orbitals.
\item {4(a,b).} The Cu and O spin relaxation rate over temperature
for a z-axis magnetic field. The O curve is only computed for
the $p_\sigma$ orbital. Including O 2s will not change the qualitative
behavior of the curve, but will increase its magnitude.
\item {5(a,b).} The Cu and O Knight shifts. The contribution from
$d_{x^2-y^2}$ is multiplied by a factor of $10$ and minus the
$d_{z^2}$ shift is plotted due to the argued sign flip arising from
interference with Cu 4s. The O shift only includes the contribution
arising from $p_\sigma$.
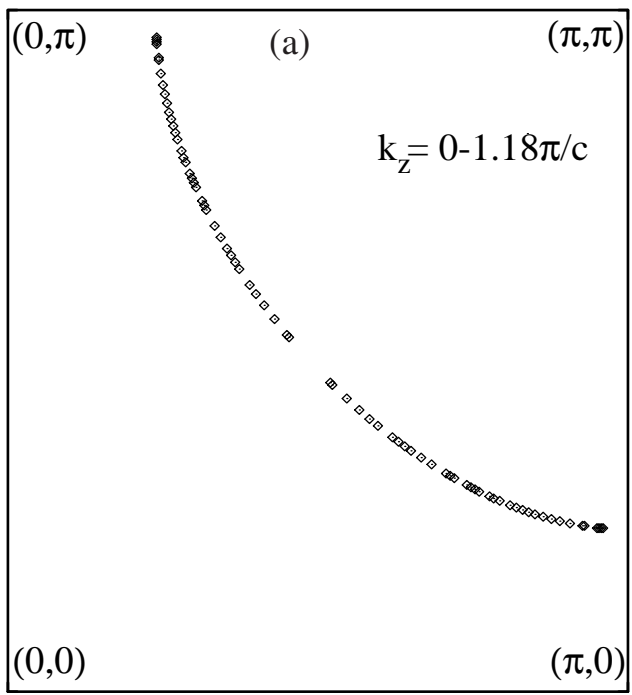
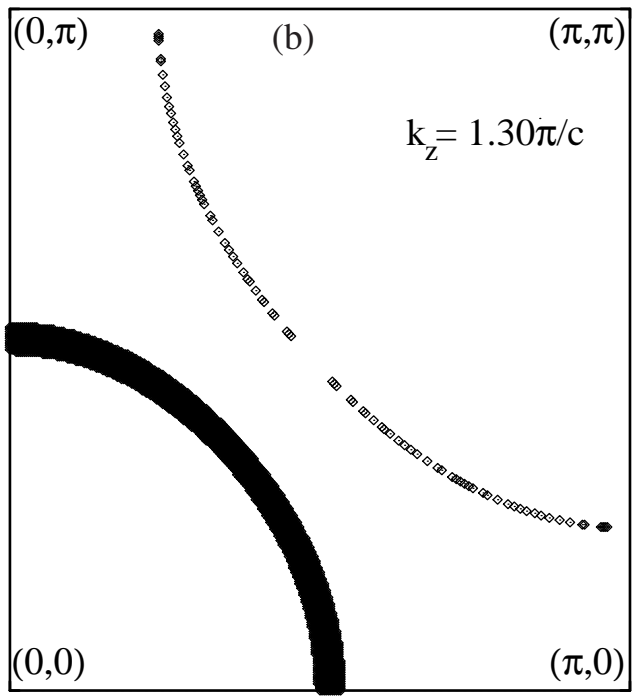
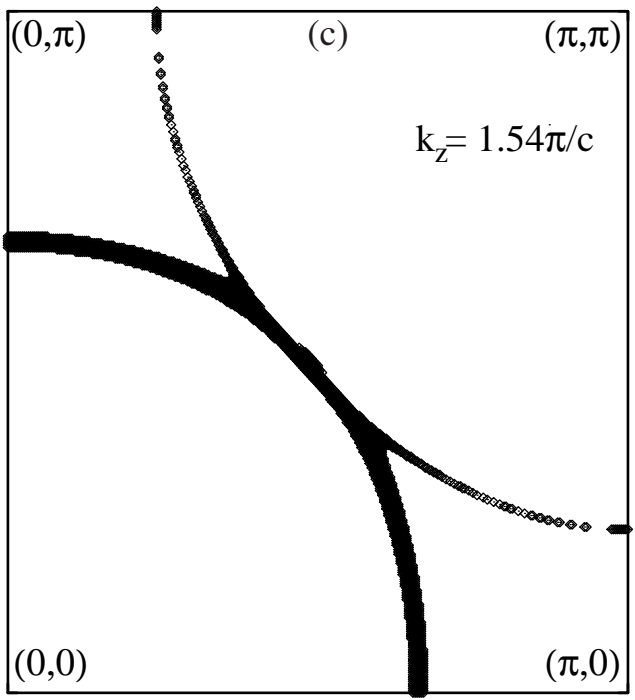
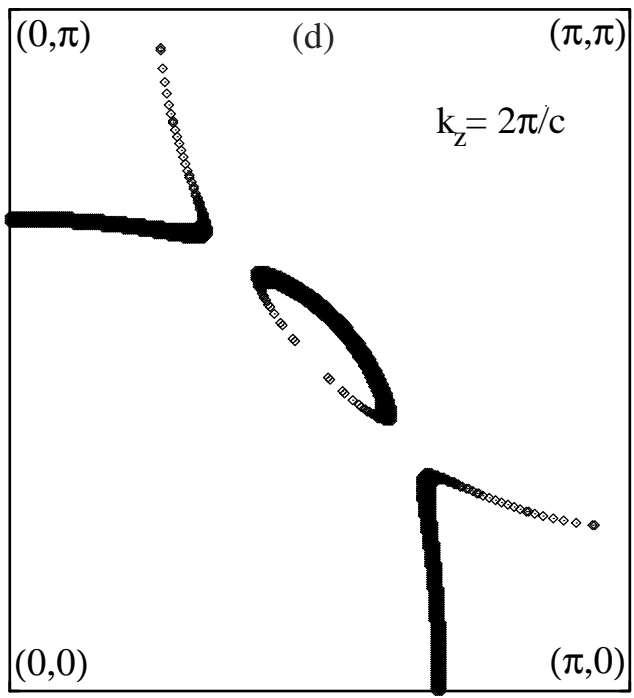
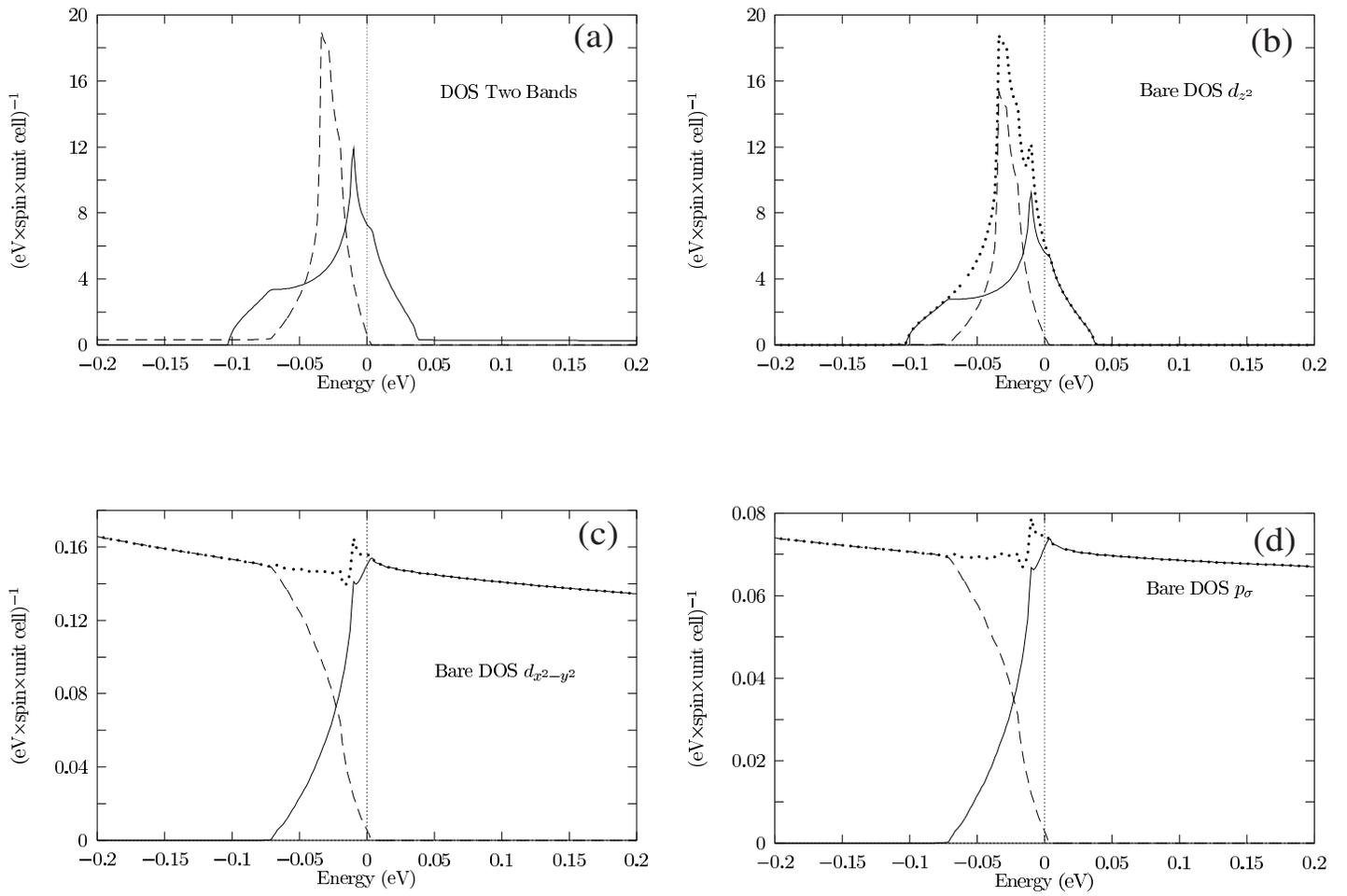
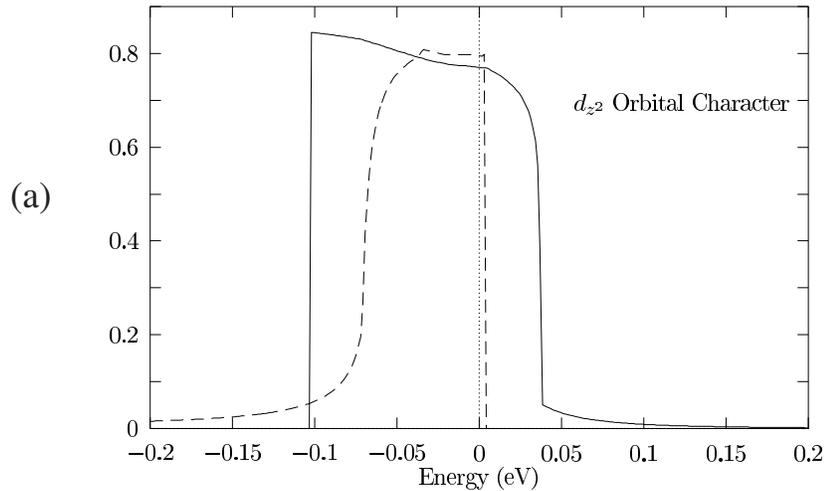
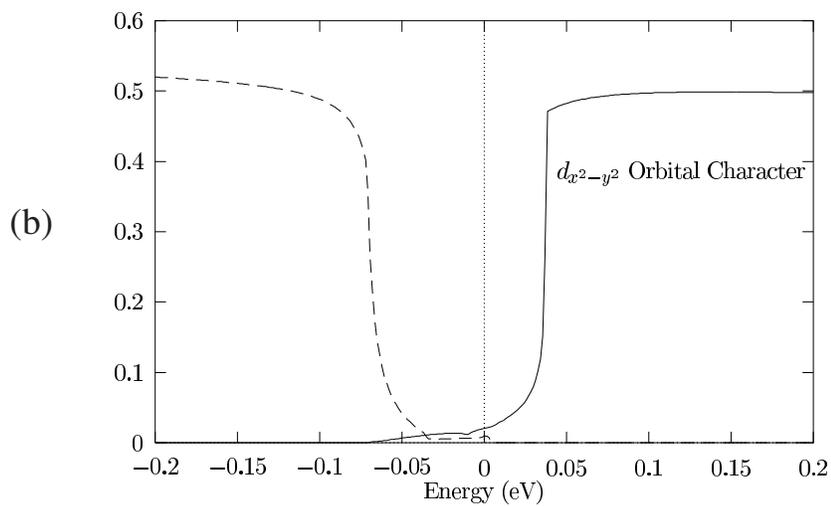
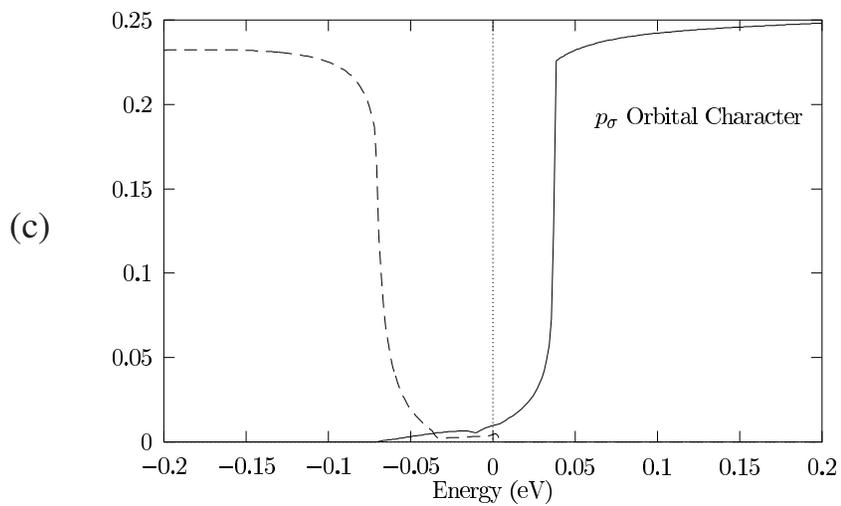
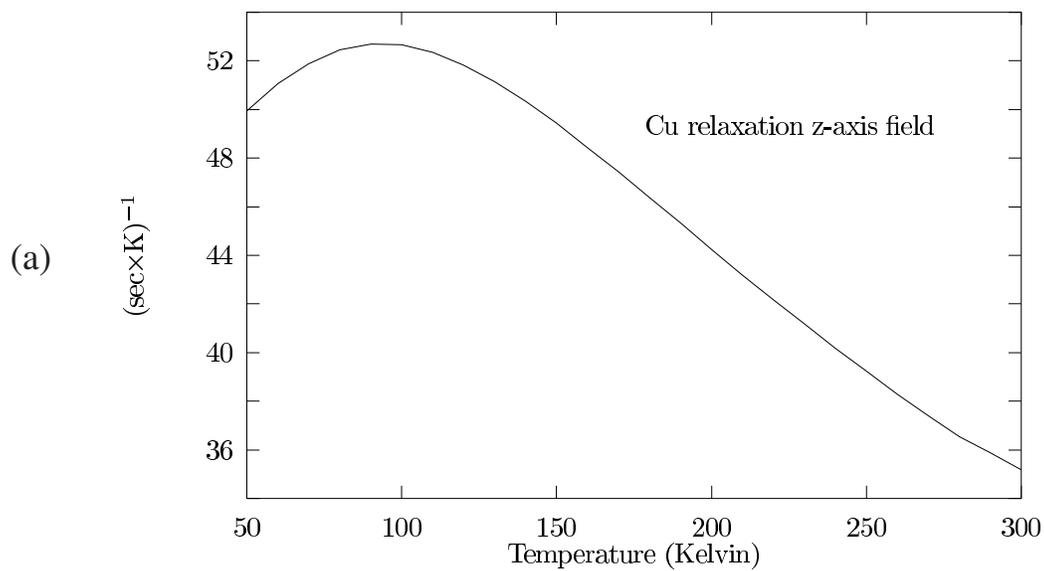
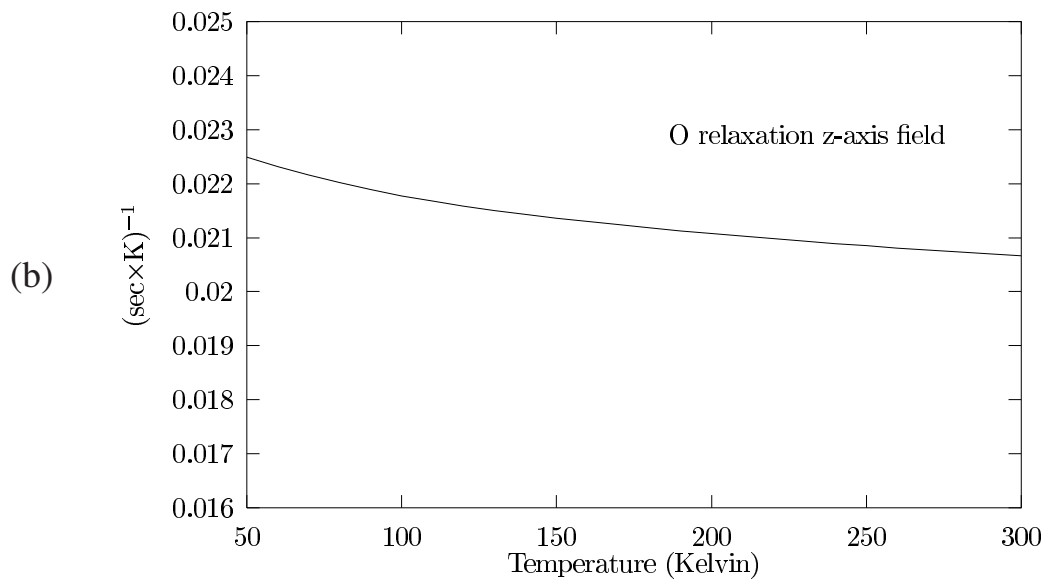
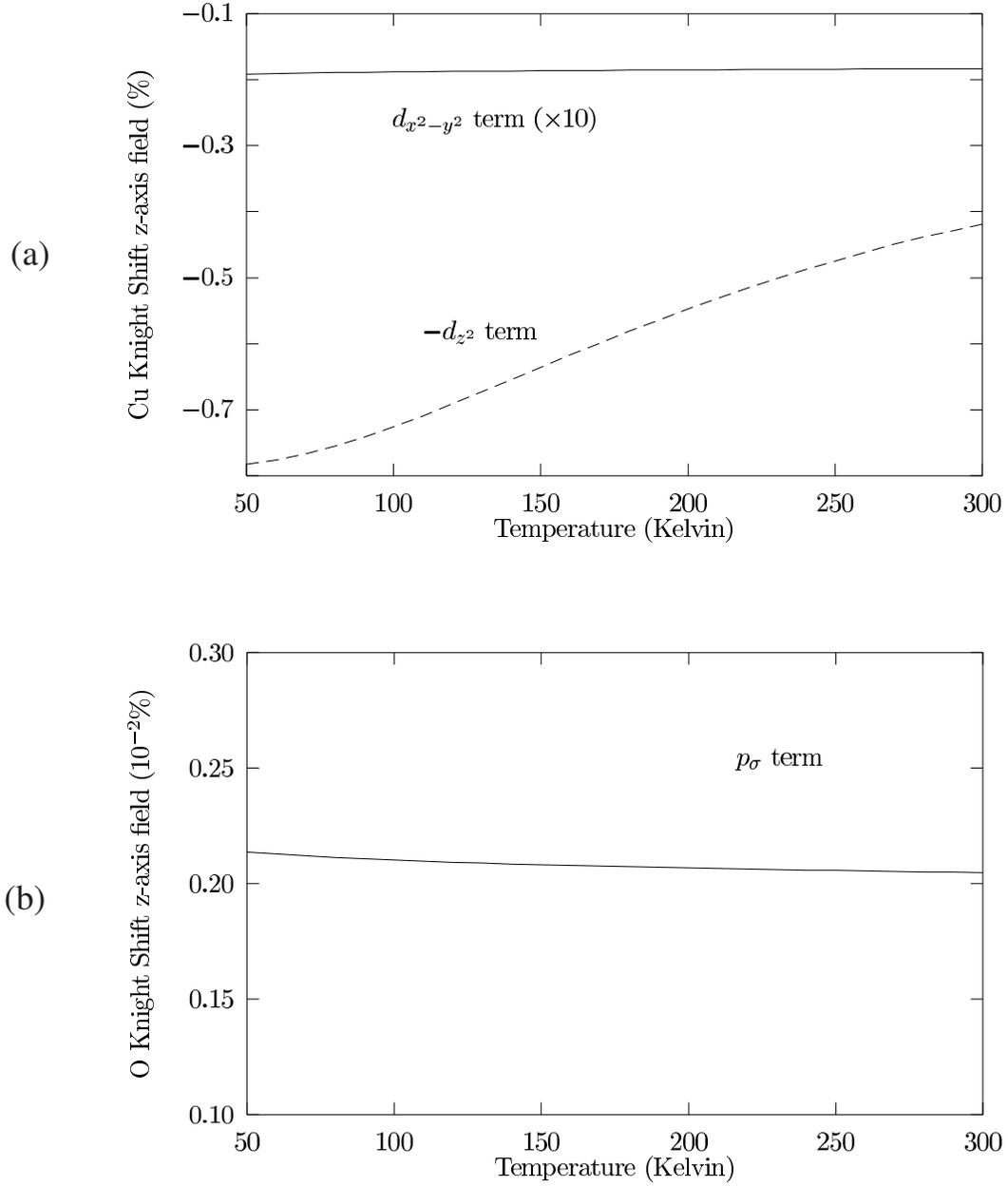
\end